\documentclass [12pt] {article}
\usepackage{amsmath}
\usepackage{amsfonts}
\newcommand{\ed}{\end{document}}

\begin{document}
\title{
Classical BRST 
charges in reducible 
BRST-anti-BRST theories} 
\author{A.~V.~Bratchikov \\ Kuban State
Technological University,\\ Krasnodar, 350072,
Russia
} 
\date {August,\,2014} 
\maketitle

\begin{abstract}
We give a solution to the classical master equation of the 
Hamiltonian BRST-anti-BRST
quantization scheme in the case of reducible gauge theories. Our approach does not 
require redefining constraints or reducibility functions. Classical BRST observables are also constructed.
\end {abstract}




\section{Introduction}
The BRST-anti-BRST symmetry in the Hamiltonian quantization of a gauge theory is generated by charges $\Omega^a,$ $a=1,2,$ satisfying the master equation 
\begin{eqnarray}
\label {1}
\{\Omega^a,\Omega^b\}=0.
\end{eqnarray} 
For irreducible theories the general solution to \eqref{1} was given in \cite{BLT1} (see also \cite{Br1}). 
Redefining constraints and reducibility functions in a  theory with linearly dependent constraints one can bring them to a 
standard 
form. 
By using this fact the existence and uniquess theorem for $\Omega^a$ in the case of reducible theories was proved \cite{BLT},\cite{GH}.

In the framework of the standard version of generalized canonical formalism, the BRST construction of  
gauge theories with linearly dependent generators was given in \cite{BF}.
The global existence of the BRST charge was proved in \cite {FHST}.
Another construction of the BRST charge for a reducible gauge theory was proposed
in \cite {Br}. This construction yields the BRST charge 
without changing constraints or reducibility functions. 

The goal of this paper is to present a 
solution to \eqref{1} in terms of the original constraints and reducibility functions 
in the case of gauge theories of any stage of reducibility. 
To discuss \eqref{1}, it is convenient to 
theat ${\Omega=(\Omega^a)}$ as an element of a Poisson algebra   
of symmetric Sp(2) tensors and to combine the Kozul-Tate differentials 
$\delta^a,$ $a=1,2,$ into a single operator $\delta.$
Our approach is based on 
a special representation of 
$\delta.$
We find a coordinate system in the configuration space 
that brings $\delta$ to a linear derivation\footnote{
Any derivation that 
leaves 
a space of linear polynomials invariant  
is called a linear derivation.}. This enables us to construct a generalized inverse of $\delta.$ Then \eqref{1} is solved by a simple
iterative procedure. We also give a solution to the equation determining the classical BRST observables.

The paper is organized as follows.
In section 2, we review the master equation,
introduce the Poisson algebra of symmetric Sp(2) tensors 
and rewrite \eqref{1} in  
terms of that algebra.
In section 3, we introduce new variables in the configuration space and find a generalized inverse of $\delta.$
A solution to the master equation 
is given in section~4. 
In section 5, a realization of the observable algebra is described .

In what follows the Grassmann parity
and new ghost number 
of a function $X$ are denoted by $\epsilon (X)$ 
and $\mbox{ngh}(X),$ 
respectively.The Poisson superbracket in phase space $\Gamma=(P_A,Q^{A}), \epsilon (P_A)=\epsilon (Q^{A}),$  
is given by 
\begin{eqnarray*}
\{X, Y\}= \frac {\partial X} {\partial Q^{A}} \frac 
{\partial Y} {\partial P_A} - (-1)^{\epsilon (X)\epsilon (Y)}\frac 
{\partial Y} {\partial Q^{A}} \frac {\partial X} {\partial P_A}.
\end{eqnarray*}  
Derivatives with respect to generalized momenta $P_A$ are always understood as left-hand, and those with respect to generalized coordinates $Q^A$ as right-hand ones.
Indices of the symplectic group Sp(2) are denoted by latin lowercase letters $a_1,a_2,\ldots.$
 For a function $X_{a_1a_2\ldots a_n}$
\begin{eqnarray*}X_{\{a_1a_2\ldots a_n\}}=X_{a_1a_2\ldots a_n}+ 
\mbox {cycl. perm.}\, 
({a_1,a_2,\ldots, a_n}).
\end{eqnarray*}

\section {Master equation for the BRST charge}
Let $
(p_i,q^i, i =1,\ldots,m)$ be the phase space coordinates, and 
let $T_{\alpha_0},$ ${\alpha_0 =1,\ldots,m_0,}$ 
$m_0\le 2m,$ be the 
first class constraints which satisfy the following Poisson brackets
\begin{eqnarray*}
\{T_{\alpha_0},T_{\beta_0}\}
=T_{\gamma_0}U_{{\alpha_0}{\beta_0}}^{\gamma_0},
\end{eqnarray*}
where $U_{{\alpha_0}{\beta_0}}^{\gamma_0}$ are phase space functions.  
The constraints are assumed to be of definite Grassmann parity $\epsilon_{\alpha_0},$ 
$\epsilon(T_{\alpha_0})=\epsilon_{\alpha_0}.$ 

We shall consider a reducible theory of $L$-th order. 
That is, there exist phase space functions 
\begin{eqnarray*}
Z^{\alpha_k}_{\alpha_{k+1}},\qquad  k=0,\ldots,L-1,\qquad \alpha_k=1,\ldots,m_k, 
\end{eqnarray*} 
such that at each stage the $Z$'s form a complete set,
\begin{eqnarray*}
Z^{\alpha_k}_{\alpha_{k+1}}\lambda^{a_{k+1}}\approx 0 \Rightarrow \lambda^{\alpha_{k+1}}\approx Z^{\alpha_{k+1}}_{\alpha_{k+2}}\lambda^{\alpha_{k+2}},\qquad k=0,\ldots,L-2,
\end{eqnarray*}
\begin{eqnarray*}
Z^{\alpha_{L-1}}_{\alpha_L}\lambda^{\alpha_L}\approx 0 \Rightarrow \lambda^{\alpha_L}\approx 0.
\end{eqnarray*}
\begin{eqnarray}\label {rc} 
T_{\alpha_0}Z^{\alpha_0}_{\alpha_1}=0,\qquad 
Z^{\alpha_{k}}_{ \alpha_{k+1}}Z^{\alpha_{k+1}}_{\alpha_{k+2}}
=T_{\beta_0}A^{\beta_0\alpha_{k}}_{\alpha_{k+2}}, \qquad  k=1,\ldots,L,
\end{eqnarray}
with
\begin{eqnarray*}
A^{\alpha_0\beta_0}_{\alpha_2}=
-(-1)^{\epsilon_{
\alpha_0}\epsilon_{\beta_0}}
A^{\beta_0\alpha_0}_{\alpha_2}.
\end{eqnarray*}
The weak equality $\approx$ means equality on the constraint surface 
\begin{eqnarray*}\Sigma:\qquad T_{\alpha_0}=0.
\end{eqnarray*}

An extended phase space of the theory 
under consideration is parametrized by the canonical variables 
\begin{eqnarray*}
{\varGamma}=(P_A,Q^A)=(\xi_\alpha;P_{A'},Q^{A'}),\qquad
(\xi_\alpha)=(p_i,q^i
),
\end{eqnarray*} 
\begin{eqnarray*}
(P_{A'},Q^{A'})= (
{\cal P}_{\alpha_{s}|a_1\ldots a_{s+1}}, c^{\alpha_s|a_1\ldots a_{s+1}};
\lambda_{\alpha_s|a_1\ldots a_s},\pi^{\alpha_s|a_1\ldots a_s};
s=0,\ldots,L),
\end{eqnarray*}
\begin{eqnarray*}
{\cal P}_{\alpha_s}, 
c^{\alpha_s}\in {\mathcal S}^{s+1},
\qquad \lambda_{\alpha_s},\pi^
{\alpha_s}\in {\mathcal S}^s,
\end{eqnarray*}
\begin{eqnarray*}
\left.\lambda_{\alpha_s|a_1\ldots a_s}\right|_{s=0}\equiv\lambda_{\alpha_0} ,\qquad
\left.\pi^{\alpha_s|a_1\ldots a_s}\right|_
{s=0}\equiv\pi^{\alpha_0}.
\end{eqnarray*}
The Grassmann parities of the canonical variables are defined as follows:
\begin{eqnarray*}\epsilon(\xi_\alpha)=\epsilon_\alpha,\qquad 
\epsilon({\cal P}_{\alpha_{s}|a_1\ldots a_{s+1}})=\epsilon(c^{\alpha_{s}|a_1\ldots a_{s+1}})=\epsilon_{\alpha_s}+s+1,
\end{eqnarray*} 
\begin{eqnarray*} 
\epsilon(\lambda_{\alpha_s|a_1\ldots a_s})=\epsilon(\pi^{\alpha_s|a_1\ldots a_s})=\epsilon_{\alpha_s}+s.
\end{eqnarray*} 
Variables of the extended phase space are assigned 
new ghost numbers by the rule 
\begin{eqnarray*}
\mbox {ngh}(\xi_\alpha)=
0,
\end{eqnarray*} 
\begin{eqnarray*} 
\mbox {ngh}({\cal P}_{\alpha_{s}|a_1\ldots a_{s+1}})=\mbox {ngh}(c^{\alpha_{s}|a_1\ldots a_{s+1}})=s+1,
\end{eqnarray*} 
\begin{eqnarray*} \mbox {ngh}(\pi^{\alpha_s|a_1\ldots a_s})=-
\mbox {ngh}(\lambda_{\alpha_s|a_1\ldots a_s})=s+2.
\end{eqnarray*}

Eq. (\ref{1}) is supplied by the conditions 
\begin{eqnarray} \label{o7}
\epsilon (\Omega^a)=1,\qquad
\mbox {ngh} (\Omega^a)=1.
\end{eqnarray}
We shall seek $\Omega^a$ in the following form:
\begin{eqnarray*} 
\Omega^a= \Omega^a_{1}+\Pi^a,\qquad \Pi^a= \sum_{n\geq 2}\Omega^a_{n},\qquad 
\Omega^a_{n} \sim  c^{n-m}\pi^m ,
\end{eqnarray*}
\begin{eqnarray*}
 \Omega^a_1= 
T_{\alpha_0} c^{\alpha_0|a}+ \sum_{s=1}^L  \bigl({\mathcal  P}_{\alpha_{s-1}|a_1\ldots a_s }\delta^a_{s+1}Z^{\alpha_{s-1}}_{\alpha_s}+{\mathcal M}^a_{\alpha_s|a_1\ldots a_{s+1}}\bigr)
c^{\alpha_s|a_1\ldots a_{s+1} }+
\end{eqnarray*}  
\begin{eqnarray}
\label{01}
+\sum_{s=0}^L
\bigl (
\varepsilon^{ac}  {\mathcal  P}_{\alpha_s|ca_1\ldots a_s}
-
[ s/ (s+1)]
 \lambda_{\alpha_{s-1}|a_1\ldots a_{s-1}}\delta^a_{s}Z^{\alpha_{s-1}}_{\alpha_s}
+{\mathcal N}^a_{\alpha_s|a_1\ldots a_{s}}
\bigr)\pi^{\alpha_s|a_1\ldots a_{s}},
\end{eqnarray}
where ${\mathcal N}^a_{a_k},{\mathcal M}^a_{a_k}$ are unknown functions of $
({\mathcal  P},\lambda),$
${\mathcal N}^a_{a_0}={\mathcal N}^a_{a_1}={\mathcal M}^a_{a_1}=0.$
 We assume that ${\mathcal N}^a_{a_k}$ and ${\mathcal M}^a_{a_k}$ only involves ${\mathcal  P}_{\alpha_s},\lambda_{\alpha_s}$ with $ {s\le k-2}.$
Eq. (\ref{o7}) implies 
\begin{eqnarray*}
\left.{\mathcal N}^a_{\alpha_k}\right|_
{{\mathcal  P}=\lambda=0}=0 ,\qquad 
\left.{\mathcal M}^a_{\alpha_k}\right|_{{\mathcal  P}=\lambda=0}=0,
\qquad 
\left.\Pi^a\right|_{{\mathcal  P}=\lambda=0}=0.
\end{eqnarray*}

${\Omega=(\Omega^a)}$ can be treated as an element of a Poisson algebra. Let ${\mathcal S}^0$ denote the space of smooth phase space functions, ${\mathcal S}^0=C^\infty(\varGamma)$, and let ${\mathcal S}^n,$ $n \geq 1,$ be the space of the functions $X^{a_1\ldots a_n}
\in C^\infty(\varGamma)$ that are symmetric under permutation of any indices.
Given $X\in {\cal S}^q$ and $ Y\in {\cal S}^p,$ the symmetric product $X\circ Y$ is defined by
\begin{eqnarray}
\label {o99} 
(X\circ Y)^{a_1\ldots a_{n}}=\frac 1 {n!}
 \sum_{\sigma \in {\mathfrak S}_n}
X^{a_{\sigma (1)}\ldots a_{\sigma (q)}}Y^{a_{\sigma (q+1)}\ldots a_{\sigma (n)}},
\end{eqnarray}
where $n=p+q,$ and the summation is extended over the symmetric group of permutations of the numbers $1,\ldots,n.$ 
This product 
is supercommutative and  
associative \cite{KM}:
\begin{eqnarray*}
 X\circ Y=(-1)^{\epsilon(X)\epsilon(Y)}Y\circ 
X,\qquad  X\circ (Y\circ Z)=(X\circ Y)\circ Z.
\end{eqnarray*}

For any $X\in {\mathcal S}^q$, $Y\in {\cal S}^p,$ 
we define the bracket ${[.\,,.]:{\cal S}^q\times {\cal S}^p\to {\cal S}^{q+p}}$ as
\begin{eqnarray}
\label {o100}
[X, Y]= \frac {\partial X} {\partial Q^{A}}\circ \frac 
{\partial Y} {\partial P_A} - (-1)^{\epsilon (X)\epsilon (Y)}\frac 
{\partial Y} {\partial Q^{A}}\circ \frac {\partial X} {\partial P_A}.
\end{eqnarray} 
Let
${\cal S}= \bigoplus
_{q=0}^\infty {\cal S}^q.$
Products \eqref{o99} and \eqref{o100} transform ${\mathcal S}
$ into a graded Poisson algebra.
One can directly verify that
\begin{eqnarray*}
[X,Y]=-(-1)^{\epsilon(X)\epsilon(Y)}[Y,X],
\end{eqnarray*}
\begin{eqnarray*}
[X,Y\circ Z]=[X,Y]\circ Z+(-1)^{\epsilon(X)\epsilon(Y)}Y\circ [X,Z],
\end{eqnarray*}
\begin{eqnarray*}
(-1)^{\epsilon(X)\epsilon(Z)}[X,[Y,Z]] +(-1)^{\epsilon(Y)\epsilon(X)}[Y,[Z,X]]+(-1)^{\epsilon(Z)\epsilon(Y)}[Z,[X,Y]]=0,
\end{eqnarray*}
$ X,Y, Z\in {\cal S}.$ 

Let us define 
\begin{eqnarray*}
{\mathcal V}^q=\{X\in {\cal S}^q:\left.X\right|_
{T={\mathcal  P}=\lambda=0
}=0\}
\end{eqnarray*}
It is easily verified that ${\mathcal V}=\bigoplus_{q=1}^\infty {\mathcal V}^q$ is a  Poisson subalgebra of ${\cal S},$ 
and $\Omega=(\Omega^a)\in {\mathcal V}^1.$ 

The bracket $\{.\,,.\}$ splits as 
\begin{eqnarray*} 
\{X,Y\}
=
\{X,Y\}_{\xi}+ \{X,Y\}_\diamond -(-1)^{\epsilon (X)\epsilon (Y)}\{Y,X\}_\diamond,
\end{eqnarray*}
where $\{.\,,.\}_{\xi}$ refers to the Poisson bracket in the original phase space and  
\begin{eqnarray*} 
\{X,Y\}_\diamond=
\frac {\partial X} {\partial Q^{A'}} \frac 
{\partial Y} {\partial P_{A'}}.
\end{eqnarray*} 
 
Let $\delta^a
:{\mathcal S}^0\to {\mathcal S}^1
$ be defined by 
\begin{eqnarray}
\label{delta} 
\delta^a =\{\Omega^a_{1},\,.\,\}_\diamond.
\end{eqnarray}
Substituting (\ref {01}) in (\ref {1}), we get 
\begin{eqnarray} \label{ruu}
\delta^a\Omega^b_{1}+\delta^b\Omega^a_{1}=0,
\end{eqnarray}
\begin{eqnarray} 
\label{r222}
\delta^{\{a} \Pi^{b\}}+ F^{ab} + A^{\{a}\Pi^{b\}} + \{\Pi^a,\Pi^b\} =0,
\end{eqnarray}
where
$F^{ab}=\{\Omega^a_{1}, \Omega^b_{1} \}_{(p,q)},$
and the operator $A^a
$ is given by
\begin{eqnarray*}
A^aX= \{\Omega^a_{1},X \}_{\xi}-(-1)^{\epsilon(X)}\{X,\Omega^a_{1}\}_\diamond,\qquad X\in {\mathcal S}^0.
\end{eqnarray*}
Eq. (\ref{ruu}) can be written in the form
\begin{eqnarray} 
\label{m1}
\delta^a\delta^b+\delta^b\delta^a=0.
\end{eqnarray}

Given a pair of operators $u^a,$ $a=1,2,$ we define an operator ${ u:{\mathcal S}^n\to {\mathcal S}^{n+1}}$ by
\begin{eqnarray*}
( u X)^{a}= u^{a}X,\qquad n=0,\qquad
\end{eqnarray*} 
\begin{eqnarray} \label {or1}
 ( u X)^{a_1\ldots a_{n+1}}=\frac 1 {n+1}
 u^{\{a_1}X^{a_2\ldots a_{n+1}\}},
\qquad n\geq 1.
\end{eqnarray} 
Using 
\eqref{or1} we rewrite  \eqref{ruu} and
\eqref{r222} as
\begin{eqnarray}
\label{meq1}
\delta \Omega_{1}=0,
\end{eqnarray}
\begin{eqnarray}
\label{meq2}
\delta\Pi+F+  A\Pi+ \frac 1 2 [\Pi,\Pi]  =0,
\end{eqnarray}
where $\Omega_{1}=(\Omega_{1}^a),$ $\Pi=(\Pi^a),$ $F=(F^{ab}).$ 
Eq. \eqref{meq1} expresses 
the nilpotency of $\delta:$
\begin{eqnarray*} 
\delta^2=0.
\end{eqnarray*}

Let us define
\begin{eqnarray*}
{
W}^a=\sum_{s=1}^L
\bigl({\mathcal M}^a_{\alpha_s|a_1\ldots a_{s+1}}
c^{\alpha_s|a_1\ldots a_{s+1}}+{\mathcal N}^a_{\alpha_s|a_1\ldots a_{s}}\pi^{\alpha_s|a_1\ldots a_{s}}\bigr),
\end{eqnarray*} 
\begin{eqnarray*} 
Q^{ab}=
2\sum_{k=2}^{L} \Bigl( {\cal P}_{\alpha_{k-2}|a_1\ldots a_{k-1}} Z^{\alpha_{k-2}}_{\alpha_{k-1}}
Z^{\alpha_{k-1}}_{\alpha_k}c^{\alpha_k|aba_1\ldots a_{k-1}}+
\end{eqnarray*}
\begin{eqnarray*} 
+\frac {(k-1)} {k+1}
\lambda_{\alpha_{k-2}|a_1\ldots a_{k-2}}Z^{\alpha_{k-2}}_{\alpha_{k-1}}
Z^{\alpha_{k-1}}_{\alpha_k}
\pi^{\alpha_s|aba_1\ldots a_{s-2}}\Bigr),
\end{eqnarray*}
and let 
\begin{eqnarray*} 
B^aX=\sum_{k=1}^{L} \Bigl(  \frac{\partial X }{\partial c^{a_{k-1}|a_1\ldots a_k}}
Z^{\alpha_{k-1}}_{\alpha_k}c^{\alpha_k|aa_1\ldots a_{k}}+ 
\end{eqnarray*}
\begin{eqnarray*} 
+ \bigl( \varepsilon^{ab} \frac{\partial X }{\partial c^{\alpha_{k}|ba_1\ldots a_{k}}}
+   \frac{k }{k+1} \frac{\partial X }{\partial \pi^{\alpha_{k-1}|a_1\ldots a_{k-1}}}
Z^{\alpha_{k-1}}_{\alpha_k}\delta^a_{a_k}
\bigr)\pi^{\alpha_{k}|a_1\ldots a_{k}}\Bigr),\qquad X\in {\mathcal S}^0. 
\end{eqnarray*}
Then (\ref{meq1}) becomes
\begin{eqnarray} 
\label {meq10}
\delta {
W}+Q+ B{
W}=0.
\end{eqnarray}
For $L=2$  (\ref{meq10}) is satisfied by ${\mathcal M}^a_{\alpha_1}={\mathcal N}^a_{\alpha_1}=0,$
\begin{eqnarray*}
{\mathcal M}^a_{\alpha_2|a_1a_2a_3}=\frac {1} {6}\left(
{\cal P}_{\alpha_{0}| a_1}{\cal P}_{\beta_{0}| a_2}\delta^a_{a_3}+ {\rm cycl. perm.} (a_1,a_2,a_3)
\right)A^{\beta_0\alpha_0}_{\alpha_2}(-1)^{\epsilon_{\alpha_0}},
\end{eqnarray*}
\begin{eqnarray}
\label{en}
{\mathcal N}^a_{\alpha_2|a_1a_2}=\frac {1} {6}\lambda_{\alpha_0}
{\cal P}_{\beta_{0}|\{ a_1}\delta^a_{a_2\}}A^{\beta_0\alpha_0}_{\alpha_2}(-1)^{\epsilon_{\alpha_0}}.
\end{eqnarray}

We shall need two auxilliary equations. 
Let $L
$ denote the left-hand side of 
\eqref{meq1}, 
$$L= \delta \Omega_1=\delta {
W}+Q+ B{
W}.$$ By using the definition of $ \delta,$ we get
\begin{eqnarray} \label {ff1}
\delta L=[\Omega_{1},L]_\diamond.
\end{eqnarray}
where 
\begin{eqnarray*}
[X,Y]_\diamond=
\frac {\partial X} {\partial Q^{A'}}\circ \frac 
{\partial Y} {\partial P_{A'}},\qquad  X, Y\in {\mathcal S}.
\end{eqnarray*} 

If (\ref{meq1}) holds, then $\{\Omega^a,\Omega^b\}=R^{ab},$ where $R^{ab}$ is  the left-hand side of (\ref {meq2}), 
\begin{eqnarray*}
R^{ab}=\delta^{\{a} \Pi^{b\}}+
  F^{ab} + A^{\{a}\Pi^{b\}} + \{\Pi^a,\Pi^b\} .
\end{eqnarray*}
From the Jacobi identity 
\begin{eqnarray*} 
\{\Omega^{a},\{\Omega^{b},\Omega^{c}\}\}+\mbox {cycl. perm.}\, (a,b,c)=0
\end{eqnarray*}
it follows that $[\Omega,R]=0,$  $\Omega=(\Omega^{a}),$ $R=
(R^{ab}),$ or equivalently
\begin{eqnarray} \label{aux}
 \delta R+ AR +[\Pi,R]=0.
\end{eqnarray}
Here we have used the relation 
$$
\{\Omega^a_{1},\,.\,\}=\delta^a+A^a.$$

\section {Generalized inversion of $\delta$}
For $k=L-2,$ (\ref{rc}) reads 
\begin{eqnarray} 
\label{r1}
Z_{\alpha'_{L-1}}^{\alpha_{L-2}}Z_{\alpha_L^{\phantom {\prime}}}^{\alpha'_{L-1}}+ 
Z^{\alpha_{L-2}}_{
A_{L-1}}Z^{
A_{L-1}}_
{\alpha_L^{\phantom {\prime}}} \approx 0,
\end{eqnarray}
where
$\alpha'_{L-1},A_{L-1}$ are index sets, such that $\alpha'_{L-1}\cup\, A_{L-1}= \alpha_{L-1},$ ${|\alpha'_{L-1}|=|\alpha_L|}$ and ${\rm rank}\,  Z^{\alpha'_{L-1}}_{\alpha_L^{\phantom{\prime}}}=|\alpha_L|.$ 
For an index set $i= \{i_1,i_2,\ldots,i_n\},$ we denote $|i|=n.$
 From  (\ref{r1}) it follows
that ${\rm rank}\, Z^{\alpha_{L-2}}_{\alpha_{L-1}}=|\alpha_{L-1}|-|\alpha_L|=|A_{L-1}|,$ and  ${\rm rank}\, Z^{\alpha_{L-2}}_{A_{L-1}}=| A_{L-1}|.$

One can split the index set $\alpha_{L-2}$ as
$\alpha_{L-2}=\alpha'_{L-2}\cup\, A_{L-2},$ such that  ${|\alpha'_{L-2}|=|A_{L-1}|,}$ and ${\rm rank}\, Z^{\alpha'_{L-2}}_{A_{L-1}}=|A_{L-1}|.$
For $k=L-3,$  (\ref{rc}) implies 
\begin{eqnarray*} 
\label{r2}
Z_{\alpha'_{L-2}}^{\alpha_{L-3}}Z_{A_{L-1}}^{\alpha'_{L-2}}+Z^{\alpha_{L-3}}_{A_{L-2}}
Z^{A_{L-2}}_{A_{L-1}}\approx 0.
\end{eqnarray*}
From this it follows that 
\begin{eqnarray*}
{\rm rank}\, Z^{\alpha_{L-3}}_{A_{L-2}}=
{\rm rank}\, Z^{\alpha_{L-3}}_{\alpha_{L-2}}=|\alpha_{L-2}|-|A_{L-1}|=|A_{L-2}|.
\end{eqnarray*}

Using induction on $ k$, we obtain a set of nonsingular matrices $Z^{\alpha'_{k-1}}_{A_{k}},$
${k=2,\ldots,L,}$
and a set of matrices $Z^{\alpha_{k-1}}_{A_{k}},k=1,\ldots,L,$ such that
\begin{eqnarray*}
{\rm rank}\, Z^{\alpha_{k-1}}_{A_{k}}= Z^{\alpha_{k-1}}_{\alpha_{k}}=|A_k|.
\end{eqnarray*}
Here $\alpha'_k\cup A_k=\alpha_k,$ $k=1,\ldots,L-1,$ and $A_{L}= \alpha_L.$

Eq. (\ref{rc}) implies 
\begin{eqnarray} \label{o2}
T_{\alpha'_0}Z^{\alpha'_0}_{A_1} + T_{A_0} Z^{A_0}_{A_1}= 0,
\end{eqnarray}
where $\alpha'_0\cup A_0=\alpha_0,$  $|\alpha'_0|=| A_1|,$ ${\rm rank}\, Z^{\alpha'_0}_{A_1}= | A_1|.$
From \eqref{o2} it follows that $T_{A_0}$ are independent. 
We assume that $T_{A_0}$ satisfy the regularity conditions. It means that there are some functions $F_{\mathcal A}(\xi),$ ${{\mathcal A}\cup A_0=\alpha=(1,\ldots,2m)},$ such that $(F_{{\mathcal A}},G_{A_0})$  can be locally taken as new coordinates in the original phase space.

Let $f: A_{k+1} \to \alpha_{k},$ $k=~0,\ldots, L-~1,$ be an embedding, $ {f(A_{k+1})=A_{k+1}\in \alpha_k,}$
and let $\bar \alpha_{k}$ be defined by $\alpha_{k}= f(A_{k+1})\cup \bar \alpha_{k}.$ Since $|A_{k}|=|\bar \alpha_{k}|,$ one can write $\bar \alpha_{k}=g(A_{k})$ for some function $g,$ and consequently
$$\alpha_{k}= f(A_{k+1})\cup g(A_{k}), \qquad k=~0,\ldots, L-~1.$$

Eq. (\ref{delta}) implies
\begin{eqnarray*}
\delta^a 
{\xi}_{\alpha}=0,\qquad \delta^a 
{\cal P}_{\alpha_0|b}=T_{\alpha_0}\delta^a_b,\qquad \delta^a \lambda_{\alpha_0}=
\varepsilon^{ab} 
{\cal P}_{\alpha_0|b},
\end{eqnarray*}
\begin{eqnarray*}
\delta^a 
{\cal P}_{\alpha_s|a_1\ldots a_{s+1}}=\frac 1 {s+1} {\cal P}_{\alpha_{s-1}|\{a_1\ldots a_s}\delta^a_{a_{s+1}\}}Z^{\alpha_{s-1}}_{\alpha_s}+ {\mathcal M}^a_{\alpha_s|a_1\ldots a_{s+1}},
\end{eqnarray*}
\begin{eqnarray}
\label{var}
\delta^a \lambda
_{\alpha_s|a_1\ldots a_s}
= \varepsilon^{ab} 
{\cal P}_{\alpha_s|ba_1\ldots a_s}
-\frac 1 {s+1}
\lambda_{\alpha_{s-1}|\{a_1\ldots a_{s-1}}
\delta^a_{a_{s}\}}
Z^{\alpha_{s-1}}
_{\alpha_s}+{\mathcal N}^a_{\alpha_s|a_1\ldots a_s} ,
\end{eqnarray}
where $s=1,\ldots,L.$

We shall use the substitution  (compare with \eqref{en})
\begin{eqnarray*}
{\mathcal M}^a_{\alpha_s|a_1\ldots a_{s+1}}=\frac 1 {s+1} {\mathcal M}_{\alpha_s|\{a_1\ldots a_s}
\delta^a_{a_{s+1}\}},\qquad{\mathcal M}_{\alpha_s}\in {\mathcal S}^s,
\end{eqnarray*}
\begin{eqnarray}
\label{pu}
{\mathcal N}^a_
{\alpha_s|a_1\ldots a_s}=\frac 1 {s+1}
{\mathcal N}_{\alpha_s|\{a_1\ldots a_{s-1}}\delta^a_{a_s\}},\qquad {\mathcal N}_{\alpha_s}\in {\mathcal S}^{s-1}.
\end{eqnarray}
From \eqref{pu} it follows that
\begin{eqnarray*}
{\mathcal M}_{\alpha_s|a_1\ldots a_s}=
\frac {s+1} {s+2} {\mathcal M}^a_{\alpha_s|aa_1\ldots a_s},\qquad
{\mathcal N}_
{\alpha_s|a_1\ldots a_{s-1}}=
{\mathcal N}^a_
{\alpha_s|aa_1\ldots a_{s-1}}.
\end{eqnarray*}

Given $X_\alpha\in {\mathcal S}^{m+n},$ we denote
\begin{eqnarray*}
X_{\alpha|(m,n)}={X}_{\alpha|{ \underbrace {\scriptstyle 1\ldots 1}
_{{m}}}\underbrace {\scriptstyle 2\ldots 2}
_{n}}.
\end{eqnarray*}
Then \eqref{var} becomes
\begin{eqnarray*}
\delta^a 
{\xi}_{\alpha}=0,\qquad
\delta^1 {\mathcal  P}_{\alpha_0|(r,t)}=rT_{\alpha_0}\qquad
\delta^2 
{\cal P}_{\alpha_0|(r,t)}=tT_{\alpha_0},
\end{eqnarray*}
\begin{eqnarray*}
\delta^1 
{\mathcal  P}_{\alpha_s|(r,t)}= \frac {r} {s+1}\Bigl({\cal P}_{\alpha_{s-1}|(r-1,t)} Z^{\alpha_{s-1}}_{\alpha_s}+ 
{\mathcal M}_{\alpha_s|(r-1,t)}\Bigr),
\end{eqnarray*}
\begin{eqnarray*}
\delta^2 
{\mathcal  P}_{\alpha_s|(r,t)}
=\frac {t} {s+1}\Bigl( {\cal P}_{\alpha_{s-1}|(r,t-1)} Z^{\alpha_{s-1}}_{\alpha_s}+ 
{\mathcal M}_{\alpha_s|(r,t-1)}\Bigr),
\end{eqnarray*}
\begin{eqnarray*}
\delta^1 \lambda_{\alpha_{s'}|(r',t')}
= 
{\mathcal P}_{\alpha_{s'}|(r',t'+1)}
-\frac {r'} {s'+1}\Bigl(\lambda_{\alpha_{{s'}-1}|(r'-1,t')}Z^{\alpha_{{s'}-1}}
_{\alpha_{s'}}-{\mathcal  N}_{\alpha_{s'}|(r'-1,t')}\Bigr), 
\end{eqnarray*}
\begin{eqnarray}
\label{p1}
\delta^2 \lambda
_{\alpha_{s'}|(r',t')}
= 
-{\mathcal  P}_{\alpha_{s'}|(r'+1,t')}
-\frac {t'} {{s'}+1}\Bigl(\lambda_{\alpha_{{s'}-1}|(r',t'-1)}Z^{\alpha_{{s'}-1}}
_{\alpha_{s'}}- {\mathcal N}_{\alpha_{s'}|(r',t'-1)}\Bigr). 
\end{eqnarray}
Here $s=1,\ldots,L,$ $r+t=s+1,$ $s'=0,\ldots,L,$  $r'+t'=s'.$

Eq. \eqref{p1} implies
\begin{eqnarray*}
(t+1)\delta^1 {\mathcal  P}_{\alpha_{s+1}|(r+1,t)}= (r+1)\delta^2{\mathcal  P}_{\alpha_{s+1}|(r,t+1)},
\end{eqnarray*}
\begin{eqnarray*}
(t'+1)\bigl(\delta^1 \lambda_{\alpha_{s'+1}|(r'+1,t')}-
{\mathcal  P}_{\alpha_{s'+1}|(r'+1,t'+1)}\bigr)=
\end{eqnarray*}
\begin{eqnarray*}
=(r'+1)\bigl(\delta^2 \lambda
_{\alpha_{s'+1}|(r',t'+1)}+{\mathcal P}_{\alpha_{s'+1}|(r'+1,t'+1)}\bigr).
\end{eqnarray*}

{\it Lemma.} 
The derivations $\delta^a$ satisfying \eqref{m1} are reducible to the form 
\begin{eqnarray*}
\delta^a{\xi}'_{\alpha}=\delta^a{\mathcal  P}'_{f(A_{s+1})|(r,t)}=0,
\end{eqnarray*}
\begin{eqnarray*}
\delta^1 {\mathcal  P}'_{g(A_s)|(r,t)}=\frac {1}{t+1}{\mathcal  P}'_{f(A_s)|(r-1,t)},\qquad
\delta^2 {\mathcal  P}'_{g(A_s)|(r,t)}=\frac {1}{r+1}{\mathcal  P}'_{f(A_s)|(r,t-1)},
\end{eqnarray*}
\begin{eqnarray*}
\delta^1\lambda'_{f(A_{s'+1})|(r',t')}=-\frac {t'+1}{t'+2}{\mathcal  P}'_{f(A_{s'+1})|(r',t'+1)},
\end{eqnarray*}
\begin{eqnarray*}
\delta^2\lambda'_{f(A_{s'+1})|(r',t')}
 =\frac {r'+1}{r'+2}{\mathcal  P}'_{f(A_{s'+1})|(r'+1,t')},
\end{eqnarray*}
\begin{eqnarray*}
\delta^1 \lambda'_{g(A_{s'})|(r',t')}= \frac 1 {t'+1}\lambda'_{f(A_{s'})|(r'-1,t')}+ {\mathcal  P}'_{g(A_{s'})|(r',t'+1)},
\end{eqnarray*}
\begin{eqnarray}
\label{p12}
\delta^2 \lambda'
_{g(A_{s'})|(r',t')}=\frac 1 {r'+1}
\lambda'_{f(A_{s'})|(r',t'-1)}-{\mathcal  P}'_{g(A_{s'})|(r'+1,t')},
\end{eqnarray}
by the change of variables
$(\xi,{\mathcal  P}, \lambda)\to
(\xi',{\mathcal  P}', \lambda'),$
\begin{eqnarray*} 
\xi'_{{\mathcal A}} = F_{{\mathcal A}},\qquad \xi'_{A_0} =  T_{A_0},
\end{eqnarray*}
\begin{eqnarray*}
{\mathcal  P}'_{f(A_{s+1})|(r,t)}=(t+1)\delta^1 {\mathcal  P}_{A_{s+1}|(r+1,t)}
,\qquad
{\mathcal  P}'_{g(A_s)|(r,t)}={\mathcal  P}_{A_s|(r,t)},
\end{eqnarray*}
\begin{eqnarray*}
\lambda'_{f(A_{s'+1})|(r',t')}=
(t'+1)\bigl(\delta^1 \lambda_{A_{s'+1}|(r'+1,t')}-{\cal P}_{A_{s'+1}|(r'+1,t'+1)}\bigr), \qquad
\end{eqnarray*}
\begin{eqnarray}
\label{cha}
\lambda'_{g(A_{s'})|(r',t')} =\lambda_{A_{s'}|(r',t')},
\qquad s,s'=0,\ldots,L-1,\qquad g(A_L)=A_L.
\end{eqnarray}

To prove this statement we first observe that \eqref {cha} is solvable with respect to the original variables. 
Assume that the functions $\xi_\alpha(\xi')$ have been constructed.
Then from (\ref{cha}) it follows that
\begin{eqnarray*}
{\mathcal P}_{\alpha'_{s}|(r,t)}
=\Bigl(\frac {s+2} {(r+1)(t+1)}
{\mathcal P}'_{f(A_{s+1})} -{\mathcal  P}'_{g(A_{s})}Z^{ A_{s}}_{A_{s+1}}-{\mathcal M}^{\prime }_{A_{s+1}}\Bigr)_{|(r,t)}{(Z^{ (-1)})}^ {A_{s+1}}_{\alpha'_{s}} 
\end{eqnarray*}
\begin{eqnarray*}
\lambda_{\alpha'_{s'}|(r',t')}
=-\Bigl(\frac {s'+2} {(r'+1)(t'+1)}
\lambda'_{f(A_{s'+1})} +\lambda'_{g(A_{s'})}Z^{ A_{s'}}_{A_{s'+1}}-{\mathcal N}^{\prime }_{A_{s'+1}}\Bigr)_{|(r',t')}{(Z^{(-1)})}^ {A_{s'+1}}_{\alpha'_{s'}} 
\end{eqnarray*}
\begin{eqnarray*}
{\mathcal  P}_{A_{s}|(r,t)}={\cal P}'_{g(A_{s})|(r,t)},\qquad \lambda_{A_{s'}|(r',t')}=
\lambda'_{g(A_{s'})|(r',t')},\qquad s,s'=0,\ldots L.
\end{eqnarray*}
Here and in what follows, for any $X(\xi, {\mathcal P},\lambda,c,\pi)$ we denote by $X'$ the function
\begin{eqnarray*}
X'(\xi', {\mathcal P}',\lambda',c,\pi)=X(\xi, {\mathcal P},\lambda,c,\pi).
\end{eqnarray*}
We have shown, therefore, that the variables 
$\bigl(
\xi'_{\alpha},
{\mathcal P}'_{\alpha_s},\lambda'_{\alpha_s},s=0,\ldots,L,\bigr)
$ 
are independent.
Eq. \eqref{p12} is a straightforward consequence of \eqref{p1} and \eqref{cha}.

Let us define derivations $\sigma_a,$ $a=1,2,$ 
by
\begin{eqnarray*}
\sigma_a{\xi}'_{\alpha'}=0,\qquad \sigma_1{\xi}'_{A}=
{\mathcal  P}'_{g(A_{0})|(1,0)},\qquad \sigma_2{\xi}'_{A}=
{\mathcal  P}'_{g(A_{0})|(0,1)},
\end{eqnarray*}
\begin{eqnarray*}
\sigma_1{\mathcal  P}'_{f(A_{s+1})|(r,t)}=\frac {(t+1)}{(s+2)}\Bigl((r+1){\mathcal  P}'_{g(A_{s+1})|(r+1,t)}
-\lambda'_{f(A_{s+1})|(r,t-1)}\Bigr),
\end{eqnarray*}
\begin{eqnarray*}
\sigma_2{\mathcal  P}'_{f(A_{s+1})|(r,t)}=
\frac {(r+1)}{(s+2)}\Bigl((t+1){\mathcal  P}'_{g(A_{s+1})|(r,t+1)}
+\lambda'_{f(A_{s+1})|(r-1,t)}\Bigr),
\end{eqnarray*}
\begin{eqnarray*}
\sigma_1 {\mathcal  P}'_{g(A_s)|(r,t)}=\frac {t}{s+1}\lambda'_{g(A_s)|(r,t-1)},
\end{eqnarray*}
\begin{eqnarray*}
\sigma_2 {\mathcal P}'_{g(A_s)|(r,t)}=
-\frac {r} {s+1}\lambda'_{g(A_s)|(r-1,t)},
\end{eqnarray*}
\begin{eqnarray*}
\sigma_1\lambda'_{f(A_{s'+1})|(r',t')}=\frac {(r'+1)(t'+1)}{(s'+2)} \lambda'_{g(A_{s'+1})|(r'+1,t')},
\end{eqnarray*}
\begin{eqnarray*}
\sigma_2\lambda'_{f(A_{s'+1})|(r',t')}=\frac {(r'+1)(t'+1)}{(s'+2)} \lambda'_{g(A_{s'+1})|(r',t'+1)},
\end{eqnarray*}
\begin{eqnarray*}
\sigma_1\lambda'_{g(A_{s'})|(r',t')}=
\sigma_2\lambda'_{g(A_{s'})|(r',t')}
 =0.
\end{eqnarray*}
Let $N$ be a counting operator,\begin{eqnarray*} 
N\xi'_{A}=0,\qquad
N\xi'_{A_0}=
\xi'_{A_0},
\qquad N{\cal P}'_{f(A_{s+1})}={\cal P}'_{f(A_{s+1})},\qquad N { \cal P}'_{g(A_s)}={\cal P}'_{g(A_s)},
\end{eqnarray*}
\begin{eqnarray*} 
N { \cal P}'_{g(A_s)}={\cal P}'_{g(A_s)},
\qquad N { \lambda}'_{g(A_{s'})}={\lambda}'_{g(A_{s'})},
\qquad N{\lambda}'_{f(A_{s'+1})}={\lambda}'_{f(A_{s'+1})},
\end{eqnarray*}
and let $M=\sigma_a\delta^a. $ 
Then we have 
\begin{eqnarray*}
\sigma_a \sigma_b+ \sigma_b\sigma_a= 0,\qquad \delta^a \sigma_b+ \sigma_b\delta^a=N \delta^a_b,
\end{eqnarray*}
\begin{eqnarray*} 
N\delta^a=\delta^aN , \qquad N\sigma_a=\sigma_a N,
\end{eqnarray*}
\begin{eqnarray*} 
M^2\delta^a=NM\delta^a,\qquad
\sigma_a M^2= N\sigma_a M,
\end{eqnarray*}
\begin{eqnarray} 
\label{us4} 
M^n=(2^{n-1}-1)N^{n-2}M^2-(2^{n-1}-2)N^{n-1}M, \qquad n\geq 3. 
\end{eqnarray}

With respect to the new coordinate system the condition $X\in {\mathcal V}$ 
becomes $$\left. X \right|_{\xi'_{A_0}={\cal P}'=\lambda'=0}=0.$$
The space ${\mathcal V}$ splits as 
\begin{eqnarray} 
\label {spl}
{\mathcal V}= \bigoplus_{k\geq 1} {\mathcal V}_{k}
\end{eqnarray} 
with  $
NX=kX$ for $X\in {\mathcal V}_{k}.$ Hence the operator $N
$ is invertible.

Let 
 $ 
\sigma: {\mathcal S}^{n}\to {\mathcal S}^{n-1}$ be defined by 
\begin{eqnarray*} 
 \sigma X=0,
\qquad n=0,
 \qquad 
( \sigma X)^{a_1\ldots a_{n}}=\sigma_{a}X^{a_1\ldots a_{n} a},\qquad n\geq 1.
\end{eqnarray*} 
One can directly verify that
\begin{eqnarray*} 
\sigma^2=0, \qquad  \sigma M=(M-N) \sigma,\qquad
\delta M=(M+N)\delta,
\end{eqnarray*} 
\begin{eqnarray} \label {oor}
(\sigma\, \delta+\delta\,\sigma)X=(nN+M)X, \qquad X\in {\mathcal S}^n.
\end{eqnarray} 
By using (\ref{us4}), we get  
\begin{eqnarray*} 
(nN+M)^{-1}=\frac 1 n N^{-1}-\frac {1} {n(n+2)(n+1)}((n+3)MN^{-2}-  M^{2}N^{-3}),\quad n \geq 1.
\end{eqnarray*}

Let $U: {\mathcal S}^{n}\to {\mathcal S}^{n}$ 
be defined by 
\begin{eqnarray*}U =\frac 1 6 ({11}N^{-1}- 6 MN^{-2}+ M^2 N^{-3}),\qquad n=0,\qquad
\end{eqnarray*}
\begin{eqnarray*}
U =(nN+M)^{-1},\qquad n\geq 1.
\end{eqnarray*}
Then $\delta^{+}=U\sigma $ is a generalized inverse of $\delta$
\begin{eqnarray} \label {or71}
\delta\,\delta^+\,\delta=\delta. 
\end{eqnarray} 
From  (\ref{oor}) it follows that \begin{eqnarray*} 
(\delta^+)^2=0, 
\end{eqnarray*} 
and for any $X\in {\mathcal S}^n,$ $n \geq 1,$
\begin{eqnarray} \label {dor1}
X= \delta^+\delta X + \delta \Lambda \delta^+X. 
\end{eqnarray} 
Here 
\begin{eqnarray*} 
\Lambda =\frac {1} {n(n+2)(n+1)}\left(n(n^2+4n+6)I-(n-4)MN^{-1}-2M^2N^{-2}\right),
\end{eqnarray*}
and $I$ is the identity map.

\section {Solution of the master equation}
{\bf Lowest order.} 
Substitution 
$(\xi,{\mathcal P},\lambda) \to
(\xi',{\mathcal P}',\lambda')$ in (\ref {meq10}) 
yields
\begin{eqnarray} \label{f1}
\delta {
W}'+Q'+ B'{
W}' =0.
\end{eqnarray}
Applying $\delta\delta^+$ to (\ref{f1}) and using (\ref{or71}) we get 
$$\delta  {
W}'+\delta\delta^{+}( B' {
W}'+Q')=0,$$ and consequently  
\begin {eqnarray} \label{tor}
 {
W}'+\delta^{+}( B' {
W}'+Q')= Y' ,
\end {eqnarray}  
where
\begin {eqnarray*}
Y'\in {V^0}, \qquad
\delta Y'=0,\qquad \epsilon(Y')= 1,\qquad \mbox{\rm ngh} (Y')= 1.
\end {eqnarray*}
Solving (\ref{tor}), we get  
\begin {eqnarray} \label{solu}
{
W}'=(I+\delta^{+} B')^{(-1)}(Y'-\delta^{+}Q') ,
\end {eqnarray}  
where 
\begin{eqnarray*}
(I+\delta^+ B')^{(-1)}=\sum_{m\geq 0}(-1)^m(\delta^+ B')^m.
\end{eqnarray*}

It remains to show that (\ref {solu}) satisfies (\ref{f1}). We shall use the 
approach of \cite{F}. With respect to the new coordinate system (\ref{ff1}) becomes  
\begin{eqnarray} \label {fy}
\delta L'= \{ \Omega^{\prime(1)},L'\}'_\diamond,
\end{eqnarray}
where
\begin{eqnarray*} 
L'=\delta {
W}'+ B'{
W}'+Q'.
\end{eqnarray*}
If ${
W}'$ is a solution to (\ref{tor}), then 
\begin {eqnarray*} 
 \delta^{+}{
W}'= \delta^{+}Y',
\end {eqnarray*}  
and hence
\begin{eqnarray} 
\label {fi}
 \delta^+L'= \delta^+ \delta {
W}'+\delta^+({
W}'+Q')=0.
\end{eqnarray}
Consider \eqref{fy} with condition \eqref{fi}.
Applying $\delta^+$ to (\ref{fy}), 
we get 
\begin {eqnarray*}
 L'=\delta^{+}\{\Omega^{\prime(1)},L'\}'_
\diamond,
\end {eqnarray*}
from which by iterations it follows that $L'=0.$ 

The functions ${\mathcal M}^a_{a_s},{\mathcal N}^a_{a_s}, s=1,\ldots,L, $ are found by substituting (\ref{cha}) in \begin{eqnarray*}
{\mathcal M}^{a}_{a_s
}(\xi, {\mathcal P}_{a_0},\ldots,{\mathcal P}_{a_{s-2}},\lambda_{\alpha_{0}},\ldots,\lambda_{\alpha_{s-2}})=
{\mathcal M}^{\prime a}_{\alpha_s
}(\xi', {\mathcal P}'_{a_0},\ldots, {\mathcal P}'_{\alpha_{s-2}},\lambda'_{\alpha_{0}},\ldots,\lambda'_{\alpha_{s-2}}),
\end{eqnarray*}
\begin{eqnarray*}
{\mathcal N}^{a}_{a_{s}
}(\xi, {\mathcal P}_{a_0},\ldots,{\mathcal P}_{a_{s-2}},\lambda_{\alpha_{0}},\ldots,\lambda_{\alpha_{s-2}})={\mathcal N}^{\prime a}_{a_{s}
}(\xi', {\mathcal P}'_{a_0},\ldots, {\mathcal P}'_{a_{s-2}},\lambda'_{\alpha_{0}},\ldots,\lambda'_{\alpha_{s-2}}),
\end{eqnarray*}
where  
\begin{eqnarray*} 
{\mathcal M}^{\prime a}_{a_{s}|(r,s+1-r)}=\frac {r!(s+1-r)!} {(s+1)!} \frac{\partial {
W}^{\prime a} }{\partial c^{a_s|(r,s+1-r)}},\qquad 
\end{eqnarray*}
\begin{eqnarray*} 
{\mathcal N}^{\prime a}_{a_{s}|(r,s-r)}=\frac {r!(s-r)!} {s!} \frac{\partial {
W}^{\prime a} }{\partial \pi^{a_s|(r,s-r)}}. 
\end{eqnarray*}
Here we have used the relation
\begin{eqnarray*}
W^{\prime a}=\sum_{s=1}^L\Bigl(\sum_{r=0}^{s+1}\frac {(s+1)!}{r!(s+1-r)!}
{\mathcal M}^{\prime a}_{\alpha_s|(r,s+1-r)}
c^{\alpha_s|(r,s+1-r)}+
\end{eqnarray*}
\begin{eqnarray*} 
+\sum_{r=0}^{s}\frac {s!}{r!(s-r)!}
{\mathcal N}^{\prime a}_{\alpha_s|(r,s-r)}\pi^{\alpha_s|(r,s-r)}\Bigr).
\end{eqnarray*}
Assume that ${\mathcal M}^a_{a_s},$ ${\mathcal N}^a_{a_s},$
$s\le k,$ have been constructed. It follows from (\ref{p1}) and (\ref{cha}) that 
the variables $(\xi', {\mathcal P}'_{a_0},\ldots, {\mathcal P}'_{a_{k-1}},\lambda'_{\alpha_{0}},\ldots,\lambda'_{\alpha_{k-1}})$ depend only on the functions 
${\mathcal M}^a_{a_s},$ ${\mathcal N}^a_{a_s}$ with $s\le k,$ and therefore $ {\mathcal M}^a
_{a_{k+1}},$ ${\mathcal N}^a_{a_{k+1}}$ are easily computed.
The functions ${\mathcal N}^a_{a_{k+1}},{\mathcal M}^a_{a_{k+1}}$ only involves ${\mathcal  P}_{\alpha_s},\lambda_{\alpha_s}, {s\le k-1},$ in agreement with the above assumption. 

{\bf Higher orders.} 
In the coordinate system $(\xi',{\mathcal P}',c,\lambda',\pi)$ 
\eqref{meq2} becomes 
\begin{eqnarray}
\label{meq2p}
\delta\Pi'+F'+  A\Pi'+ \frac 1 {2}[\Pi',\Pi']'  =0.
\end{eqnarray}
Applying 
the operator $ \delta \delta^+$ to (\ref {meq2p}), we get 
 \begin{eqnarray}
\label{d3}
 \delta\Pi'+ \delta \delta^+(F'+  A\Pi'+ \frac 1 {2}[\Pi',\Pi']') =0.
\end{eqnarray}
From (\ref{d3}) it follows that 
\begin{eqnarray}
\label{seo} \Pi'=\Upsilon- \delta^+(F'+  A\Pi'+ \frac 1 {2}[\Pi',\Pi']'),
\end{eqnarray}
where 
\begin{eqnarray*} 
\Upsilon\in V^1, \qquad  \delta \Upsilon=0, \qquad
\Upsilon= \sum_{n\geq 2}\Upsilon^{(n)},\qquad  
\Upsilon^{(n)} \sim  c
^{n-m}
\pi^m.
\end{eqnarray*}
If $\Pi'$ is a solution to (\ref{seo}) then
\begin{eqnarray}\label{see}
 \delta^+\Pi'=  \delta^+\Upsilon.
\end{eqnarray}

Let us show that a solution to (\ref{seo}) satisfies (\ref {meq2p}). Changing variables in (\ref {aux}) from $(\xi,{\mathcal P},\lambda)$ to
$(\xi',{\mathcal P}',\lambda'),$
we get 
\begin{eqnarray} \label{aux0}
 \delta R'+ AR' +[\Pi',R']'=0.
\end{eqnarray}
Consider (\ref {aux0}), where $\Pi'$ is a solution to (\ref{seo}), with the boundary 
condition 
\begin{eqnarray}
\label{au23}
 \delta^+R'=0.
\end{eqnarray}
Applying $ \delta^+$ to (\ref {aux0}),
we get
\begin{eqnarray*}
R'+ \delta^+( AR' +[\Pi',R']')=0.
\end{eqnarray*}
From this by iterations it follows that $R'=0.$
For checking (\ref{au23}) we have 
\begin{eqnarray*}
 \delta^+R'= \delta^+\delta\Pi'+ \delta^+(F'+  A\Pi'+ \frac 1 {2}[\Pi',\Pi']')= \delta^+ \delta\Pi'+\Upsilon-\Pi',
\end{eqnarray*}
and therefore by (\ref{dor1}) and (\ref{see}), $ \delta^+R=0.$

Let $\langle .\,,. \rangle:  S^1\times S^1 \to S^1$  be defined by
\begin{eqnarray} \label {obosr}
\langle X_1,X_2 \rangle = -\frac 1 2 (I+ \delta^+A)^{-1} \delta^+\left([ X_1,X_2]+[X_2,X_1] \right),
\end{eqnarray} 
where  
\begin{eqnarray*} (I+ \delta^+ A)^{-1}=\sum_{m\geq 0}(-1)^m( \delta^+ A)^m. 
\end{eqnarray*}
Then we have
\begin{eqnarray}
\label{dursd}
\Pi'= \Pi_0+\frac 1 2 \langle \Pi',\Pi' \rangle,
\end{eqnarray}
where 
\begin{eqnarray*}
\Pi_0 = (I+ \delta^+ A)^{-1}\left(\Upsilon -  \delta^+F'\right).
\end{eqnarray*} 
Iterating  (\ref {dursd}) we can construct $\Pi'$ and, consequently, the BRST charge $\Omega.$
The first two terms are
 \begin{eqnarray*} 
\Pi' = \Pi_0 + \frac 1 2 \langle \Pi_0,\Pi_0 \rangle+ \ldots.\end{eqnarray*}

\section {
Observables 
} 
Let $P$ denote the Poisson algebra of first class functions,
\begin{eqnarray*}
P =\{f(\xi)\,|\, 
\{f,T_\alpha \}\approx 0
\},
\end{eqnarray*}
and let 
\begin{eqnarray*}
J =\{u(\xi)\,|\,
u
\approx 0
\}.
\end{eqnarray*}
Elements of the Poisson algebra $P/J$ are called classical observables. 

BRST observables are determined by solutions to the equation
\begin{eqnarray}
\label {44}
[\Omega,\Phi]=0,\qquad \Phi\in S^0.
\end{eqnarray}
With respect to 
the variables $(\xi',{\mathcal P}',c,\lambda',\pi)$ (\ref{44}) becomes
\begin{eqnarray}
\label {45}
[\Omega',\Phi']'=0.
\end{eqnarray}
The boundary conditions read
\begin{eqnarray}
\label{pps2}
\mbox{\rm ngh} (\Phi')=0,\qquad \left.\Phi'\right|_{{\mathcal P}'=c=\lambda'=\pi=0}=\Phi_0
,\qquad 
\bar \sigma(\Phi'-\Phi_0)=0,
\end{eqnarray}
where $\Phi_0(\xi'
) \in P,$ $\bar \sigma=\epsilon^{ab}\sigma_a\sigma_b.$ 

The function $\Phi'$ can be written as  
\begin{eqnarray} 
\label{us0} 
\Phi' = \Phi_0+K,\qquad K=\sum_{n\geq 1}\Phi^{(n)},\qquad 
\Phi^{(n)}\sim c^{n-m}\pi^m.
\end{eqnarray}
Substituting (\ref{us0}) in (\ref{45}), we get  
\begin {eqnarray}
\label{qsp2}
\delta K+ [\Omega', \Phi_0]'+ A K + [\Pi',K]'=0
\end {eqnarray}
or equivalently, 
\begin {eqnarray}
\label{qsp3}
K +  \delta ^+([\Omega, \Phi_0]'+  A K + [\Pi',K]')={\mathcal Z}, 
\end {eqnarray}
where 
\begin{eqnarray*}{\mathcal Z}\in S^0,\qquad   \delta{{\mathcal Z}}=0,\qquad  \mbox{\rm ngh} ({\mathcal Z})=0.\end{eqnarray*}

Let us denote $
\bar \delta=\epsilon_{ab}\delta^a\delta^b.
$
Then 
\begin {eqnarray*}
\bar \delta \bar \sigma-\bar \sigma\bar \delta=4N^2-2MN,
\end {eqnarray*}
from which it follows that for any $X\in {S^0}$ 
\begin {eqnarray}
\label{qsp5}
X= \frac 1 2 MN^{-1} X + \frac 1 4 \left(
\bar \delta \bar \sigma -\bar\sigma \bar \delta\right)N^{-2}X.
\end {eqnarray}
The boundary conditions (\ref{pps2}) imply $ \bar \sigma K=0,$ and therefore $ \bar \sigma {\mathcal Z}=0,$ since $\bar \sigma \delta^+=0.$
By using (\ref{qsp5}) we get ${\mathcal Z}=0.$
Solving (\ref{qsp3}) for $K$ yields
\begin {eqnarray} 
\label{q21}
K=- (I+\delta^{+}(A+ {\rm ad}\,\Pi) )^{(-1)}\delta^+[\Omega, \Phi_0].
\end {eqnarray}  
We must now show that (\ref{qsp3}) satisfies (\ref{qsp2}).

The Jacobi identities for the functions $\Omega^{\prime a},\Phi' $ read 
\begin {eqnarray}  
\label {q1}
\{\Omega^{\prime a},\{\Omega^{\prime b},\Phi'\}'\}'+
\{\Omega^{\prime b},\{\Omega^{\prime a},\Phi'\}'\}'=0.
\end {eqnarray} 
Let $G=(G^a)$ denote left-hand side of (\ref{qsp2}), 
\begin {eqnarray*}
\label{qsp4}
G=\delta K+ [\Omega', \Phi_0]'+ A K + [\Pi',K]'. 
\end {eqnarray*}
Then (\ref{q1}) becomes 
\begin {eqnarray}  
\label {l3}
\delta G+  A G+[\Pi',G]'=0.
\end {eqnarray} 
It is easily verified that if $K$ satisfies (\ref{qsp3}) then 
${\delta^{+}K = \delta^{+}\Upsilon,}$ and
\begin {eqnarray}
\label{s5}
\delta^{+}G=0.
\end {eqnarray}
Consider equation (\ref{l3}) and boundary condition (\ref{s5}).
By using (\ref{qsp5}), we get  
\begin {eqnarray*}
G=- \delta^{+}( A G +[\Pi',G]'). 
\end {eqnarray*} 
From this it follows that $G=0.$ 

We conclude that the solution to  (\ref{45}), (\ref{pps2}) is given by  
\begin {eqnarray} 
\label{q2}
\Phi'={\mathcal L}{\Phi_0},
\end {eqnarray}  
where 
\begin {eqnarray*} 
{\mathcal L}=I- (I+\delta^{+}( A+ {\rm ad}\,\Pi) )^{(-1)}\delta^{+}
[\Omega',\,.\,]'.
\end {eqnarray*}  
The operator ${\mathcal L}$ is invertible. The inverse ${\mathcal L}^{-1}$ is given by 
\begin {eqnarray*} 
{\mathcal L}^{-1}\Phi' = \left.\Phi' \right|_{{\mathcal P}'=c=\lambda'=\pi =0}.
\end {eqnarray*}  
Eq. (\ref{q2}) establishes a one-to-one correspondence between first class functions and solutions to  (\ref {45}), (\ref{pps2}).

Let ${\mathcal L}(D)$ denote the image of $D\subset P$ under the mapping ${\mathcal L}.$  
For ${\Phi'_1,\Phi'_2 \in {\mathcal L}(P)}$ 
\begin {eqnarray*} 
\left.
 \,\,\,\{\Phi'_1,\Phi'_2\}' \right|_{{\mathcal P}'=c=\lambda'=\pi =0}=
\{
\left.\Phi'_1\right|_{{\mathcal P}'=c=\lambda'=\pi=0},
\left.\Phi'_2 \right|_{{\mathcal P}'=c=\lambda'=\pi =0}\}',
\end {eqnarray*}  
\begin {eqnarray*} 
\left.
 (\Phi'_1\Phi'_2) \right|_{{\mathcal P}'=c=\lambda'=\pi=0}=
\left.\Phi'_1\right|_{{\mathcal P}'=c=\lambda'=\pi=0}
\left.\Phi'_2 \right|_{{\mathcal P}'=c=\lambda'=\pi=0}. 
\end {eqnarray*}  
This means that ${\mathcal L}(P)$ and ${P}$ are isomorphic as Poisson algebras, and 
therefore  ${\mathcal L}(P)/{\mathcal L}(J)$ gives a realization of classical observables.

\end{document}